\documentclass[11pt]{article}
\usepackage{amsfonts,amsmath}
\usepackage{graphicx}

\def\bbt{\bibitem}
\def\be{\begin{equation}}
\def\en{\end{equation}}
\def\ber{\begin{eqnarray}}
\def\enr{\end{eqnarray}}
\def\nmb{ \nonumber\\}
\def\d{\partial}

\def\rbrc{\rbrace}
\def\lbrc{\lbrace}
\def\ov{\over }
\def\tld{\tilde}

\def\Sgm{\Sigma}
\def\al{\alpha}
\def\bet{\beta}
\def\gm{\gamma}
\def\Gm{\Gamma}

\def\lm{\lambda}
\def\Lm{\Lambda}
\def\Om{\Omega}
\def\om{\omega}

\def\Tt{\Theta}

\def\dlt{\delta}

\def\be{{\bf e}}

\def\MM{\mathbb{M}}

\begin{document}
%\nopagenumbers
%\rightline{Landau Tmp.}
\vskip 2 true cm

\centerline{\bf Fermionic screenings and chiral de Rham complex}
\centerline{\bf on CY manifolds with line bundles.}

\vskip 1.5 true cm
\centerline{\bf S. E. Parkhomenko}
\centerline{Landau Institute for Theoretical Physics}
\centerline{142432 Chernogolovka, Russia}
\vskip 0.5 true cm
\centerline{spark@itp.ac.ru}
\vskip 1 true cm
\centerline{\bf Abstract}
\vskip 0.5 true cm

 We represent a generalization of Borisov's construction of chiral de Rham complex for the case of line bundle twisted chiral de Rham complex on Calabi-Yau hypersurface in projective space. We generalize the differential associated to the polytope $\Delta$ of the projective space $\mathbb{P}^{d-1}$ by allowing nonzero modes for the screening currents forming this differential. It is shown that the numbers of screening current modes define the support function of toric divisor of a line bundle on $\mathbb{P}^{d-1}$ that twists
the chiral de Rham complex on Calabi-Yau hypersurface.

\vskip 10pt

"{\it PACS: 11.25Hf; 11.25 Pm.}"


\smallskip
\vskip 10pt
\centerline{\bf 1. Introduction}
\vskip 10pt

 The Calabi-Yau manifolds with bundles appear in various types of compactifications
of string theory. The first example is heterotic string compactification on Calabi-Yau (CY) manifolds. This is currently most successful approach to the problem
of string models construction relevant to 4-dimensional particle physics. The main ingredients of the heterotic models is a Calabi-Yau three-fold and two holomorphic vector
bundles on it. In the simplest case of "standard embedding" one of the bundles is taken to be trivial while the other one coincide with the tangent bundle of the Calabi-Yau manifold.
But explicit constructions of the bundles in general case is hard to obtain.
Nevertheless, in the series of papers \cite{Monad} the monad bundles approach has been developed
for the systematic construction of a large class of vector bundles over the Calabi-Yau
manifolds defined as complete intersections in products of projective spaces. 

 Although the monad construction of \cite{Monad} is quite general, it is purely classical,
so the Gepner models \cite{Gep} (see also  \cite{BGep}) are the only known models of the quantum string compactification.
It rises the question what is the quantum version of monad bundle construction? 

 The second example is Type IIA compactification with $D$-branes wrapping
the Calabi-Yau manifold.
In this case the Chan-Paton vector bundle appears \cite{HM}, so that the similar question make sense: what object describes the quantum strings on Calabi-Yau manifold with Chan-Paton bundles?

 In this note we analyze the most simple version of these questions when
we have only line bundle on the Calabi-Yau manifold and represent a construction of vertex operator algebra starting from Calabi-Yau hypersurace in projective space
and a line bundle defined on this space. 

 Our approach is based essentially on the work of Borisov \cite{B} where
a certain sheaf of vertex operator algebras endowed with $N=2$
Virasoro superalgebra action has been constructed for each pair of dual reflexive polytopes $\Delta$ and $\Delta^{*}$ defining CY hypersurface in toric manifold $\mathbb{P}_{\Delta}$. Thus, Borisov constructed directly holomorphic sector of the CFT from toric dates of CY manifold. The main object of his construction is a set of fermionic screening currents associated to the points of that pair of polytopes. Zero modes of these currents are used to build up a differential $D_{\Delta}+D_{\Delta^{*}}$ whose cohomology calculated in some lattice vertex algebra gives the global sections of a sheaf known as chiral de Rham complex
due to \cite{MSV}. On the local sections of the chiral de Rham complex the $N=2$ Virasoro superalgebra is acting \cite{MSV}. In CY case this algebra survives the cohomology and hence, the global sections of the chiral de Rham complex can be considered as a holomorphic sector of the space of states of the $N=2$ superconformal sigma-model on the CY manifold. The question
how the costruction \cite{B} is related to the Gepner models has been clarified considerable 
in \cite{GM} and \cite{P}.

 Borsov's construction can also be generalized by allowing non-zero modes for the screening
currents forming the differential. We consider CY hypersurface in projective
space $\mathbb{P}^{d-1}$ and generalize differential of Borisov by allowing non-zero modes only for screening currents associated to the points of $\mathbb{P}^{d-1}$ polytope $\Delta$. We thus generalize the differential $D_{\Delta}$ leaving unchanged the differential $D_{\Delta^{*}}$ defining CY hypersurface. We show that the numbers of screening current modes from $D_{\Delta}$ define the support function of toric divisor \cite{Dan},
\cite{Ful} of a line bundle on $\mathbb{P}^{d-1}$. By this means, the 
chiral de Rham complex on $\mathbb{P}^{d-1}$ appears to be twisted by the line bundle.
 
 The paper is organized as follows. In Section 2 we briefly review the construction of
paper \cite{B} for the case of CY hypersurface in $\mathbb{P}^{d-1}$. In Section 3 we calculate first the cohomology with respect to the generalized differential $D_{\Delta}$ and relate them
to the sections of chiral de Rham complex twisted by the sheaf $O(N)$. To do that we find the generalized $bc\bet\gm$ system of fields generating the cohomology. The generalization appears only for the modes of vector fields operators. They are replaced by the covariant derivative operators with $U(1)$ connection. Then we define the trivialization maps of the modules generated by these
generalized $bc\bet\gm$ fields to the modules of sections of the usual chiral de Rham complex over
the affine space and find the transition functions for different trivializations. Tese turn out to be the transition functions of the $O(N)$ bundle on $\mathbb{P}^{d-1}$, where $N$ is determined
by the numbers of modes of screening currents composing the differential $D_{\Delta}$. Moreover, we establish the relation between the numbers of screening currents modes and toric divisor support function for the line bundle $O(N)$ on $\mathbb{P}^{d-1}$. The support function and trivialization maps are consistent
with the localization maps determined by the fan structure which allows to calculate the cohomology of the twisted chiral de Rham complex
by $\check{C}$ech complex of the covering. In complete analogy with \cite{B}, the second differential $D_{\Delta^{*}}$ is used to restrict the sheaf on the CY hypersurface. 

 In Section 4 we calculate the elliptic genus of the twisted chiral de Rham complex and represent it in terms of theta functions. For a torus in $\mathbb{P}^{2}$
and $K3$ in $\mathbb{P}^{3}$ we find the limit as $q\rightarrow 0$ and relate the results to the
Hodge numbers of the sheaf $O(N)$ on the torus and $K3$. Section 5 contains the concluding remarks.

\vskip 10pt
\centerline{\bf 2. Chiral de Rham complex on CY hypersurface in $\mathbb{P}^{d-1}$.}
\vskip 10pt

 In this section we review the construction \cite{B} of chiral de Rham complex
and its cohomology for the case of CY hypersurface in projective space $\mathbb{P}^{d-1}$.

 Let $\lbrc e_{1},...,e_{d}\rbrc$ be the standard basis in $\mathbb{R}^{d}$
and $\Lm\subset \mathbb{R}^{d}$ be the lattice generated by the vectors
$e_{0}=\frac{1}{d}(e_{1}+...+e_{d})$, $e_{1}$, ...,$e_{d}$:
\ber
\Lm=\mathbb{Z}e_{0}\oplus \mathbb{Z}e_{1}\oplus ...\oplus \mathbb{Z}e_{d}
\label{2.lat}
\enr
We then consider the  fan $\Sgm\subset \Lm$ \cite{Dan}, \cite{Ful} encoding the toric data of an $O(d)$-bundle 
\ber
\pi: E\rightarrow \mathbb{P}^{d-1}
\label{2.bundle}
\enr
The maximal dimension cones of the fan are d-dimensional cones 
$C_{I}\subset \Lm$, $I=1,2,...,d$, spanned by the vectors $e_{0},...,\hat{e_{I}},...,e_{d}$,
where the vector $e_{I}$ omitted. The intersection of the maximal dimension cones
is also a cone in $\Sgm$ 
\ber
C_{I}\cap C_{J}\cap...\cap C_{K}=C_{IJ...K}\in \Sgm .
\label{2.CIJK}
\enr
All faces of the cone from $\Sgm$ are the cones from $\Sgm$. (See \cite{Dan}, \cite{Ful} for more detailed definition of fan.)

 Let $\lbrc e^{*}_{1},...,e^{*}_{d}\rbrc$ be the dual basis to the standard one
$\lbrc e_{1},...,e_{d}\rbrc$ and let $\Lm^{*}$ be the dual lattice to $\Lm$. For every cone $C\in \Sgm$, one considers
the dual cone $C^{*}\in \Lm^{*}$ defined by
\ber
C^{*}=\lbrc p^{*}\in\Lm^{*}|p^{*}(C)\geq 0\rbrc
\label{2.dcone}
\enr
as well as the affine variety $A_{C}=Spec(\mathbb{C}[C^{*}])$. If $C^{*}$ is a face
of $\tld{C}^{*}$ then $\mathbb{C}[C^{*}]$ is a localization of $\mathbb{C}[\tld{C}^{*}]$
by the monomials $a^{p^{*}}\in \mathbb{C}[C^{*}]$, where $p^{*}\in \tld{C}^{*}$ and $p^{*}(C)=0$. It allows
to glue $A_{C}$ to form the space $E$.

  The polytope $\Delta$ of
$\mathbb{P}^{d-1}$ is given by the points from $\Sgm$ satisfying the
equation
\ber
deg^{*}(\Sgm)=1,
\label{2.Delta}
\enr
where 
\ber
deg^{*}=e^{*}_{1}+...+e^{*}_{d}
\label{2.degdual}
\enr

 Let $X_{i}(z), X^{*}_{i}(z)$, $i=1,2,...,d$ be free bosonic fields and $\psi_{i}(z), \psi^{*}_{i}(z)$, $i=1,2,...,d$ be 
free fermionic fields  
so that its OPE's are given by
\ber
X^{*}_{i}(z_{1})X_{j}(z_{2})=\ln(z_{12})\dlt_{i,j}+reg.,\nmb
\psi^{*}_{i}(z_{1})\psi_{j}(z_{2})=z_{12}^{-1}\dlt_{i,j}+reg,
\label{2.ope}
\enr
where $z_{12}=z_{1}-z_{2}$.

 The fields are expanded into the integer modes
\ber
\d X^{*}_{i}(z)=\sum_{n\in\mathbb{Z}}X^{*}_{i}[n]z^{-n-1},
\
\d X_{i}(z)=\sum_{n\in\mathbb{Z}}X_{i}[n]z^{-n-1},
\nmb
\psi^{*}_{i}(z)=\sum_{n\in\mathbb{Z}}\psi^{*}_{i}[n]z^{-n-\frac{1}{2}},
\
\psi_{i}(z)=\sum_{n\in\mathbb{Z}}\psi_{i}[n]z^{-n-\frac{1}{2}}
\label{2.modes}
\enr
We therefore consider  Ramond sector.

 To the lattice
${\Gm}=\Lm\oplus \Lm^{*}$  we associate the direct sum of Fock spaces
\ber
\Phi_{\Gm}=\oplus _{(p,p^{*})\in\Gm}F_{(p,p^{*})},
\label{2.Fock}
\enr
where $F_{(p,p^{*})}$ is the Fock module generated by $X_{i}[n]$, $X^{*}_{i}[n]$,
$\psi_{i}[n]$, $\psi^{*}_{i}[n]$ from the vacuum $|p,p^{*}>$ determined by
\ber
X^{*}_{i}[n]|p,p^{*}>=X_{i}[n]|p,p^{*}>=\psi_{i}[n]|p,p^{*}>=\psi^{*}_{i}[n-1]|p,p^{*}>=0, n>0,
\nmb
X^{*}_{i}[0]|p,p^{*}>=p^{*}_{i}|p,p^{*}>,
\
X_{i}[0]|p,p^{*}>=p_{i}|p,p^{*}>
\label{2.Fvac}
\enr

 For each vector $e_{i}$, $i=0,1,...,d$ generating 1-dimensional cone from $\Sgm$, we define the fermionic screening current and screening charge
\ber
S^{*}_{i}(z)=e_{i}\cdot\psi^{*}\exp (e_{i}\cdot X^{*})(z), \ 
Q^{*}_{i}=\oint dz S^{*}_{i}(z)
\label{2.chrg}
\enr
We form the BRST operators for each maximal dimension cone $C_{I}$
\ber
D^{*}_{I}=Q^{*}_{0}+...+\hat{Q^{*}}_{I}+...+Q^{*}_{d}
\label{2.dbrst}
\enr
where $Q^{*}_{I}$ is omitted. Then, one considers the space
\ber
\Phi_{C_{I}\otimes \Lm^{*}}=\oplus _{(p,p^{*})\in C_{I}\otimes \Lm^{*}}F_{(p,p^{*})}
\label{2.iFock}
\enr
The space of sections $M_{C_{I}}$ of the chiral de Rham complex over the $A_{C_{I}}$ is given by the cohomology of $\Phi_{C_{I}\otimes \Lm^{*}}$ with respect to the operator $D^{*}_{I}$.
It is generated by the following fields ~\cite{B}
\ber
a_{I\mu}(z)=\exp{[w^{*}_{I\mu}\cdot X]}(z),\ \al_{I\mu}(z)=w^{*}_{I\mu}\cdot\psi\exp{[w^{*}_{I\mu}\cdot X]}(z),
\nmb
a^{*}_{I\mu}(z)=(e_{\mu}\cdot\d X^{*}-w^{*}_{I\mu}\cdot\psi_{i}e_{\mu}\cdot\psi^{*}_{i})
\exp{[-w^{*}_{I\mu}\cdot X]}(z), \
\al^{*}_{I\mu}(z)=e_{\mu}\cdot\psi^{*}\exp{[-w^{*}_{I\mu}\cdot X]}(z)
\label{2.btgm}
\enr
where $w^{*}_{I\mu}$ are the dual vectors to the basis of vectors $\left\{e_{\mu},\mu=0,...,\hat{I},...d\right\}$ generating the cone $C_{I}$:
\ber
<w^{*}_{I\mu},e_{\nu}>=\dlt_{\mu \nu}
\label{2.edual}
\enr

The singular operator product expansions of these fields are
\ber
a^{*}_{I\mu}(z_{1})a_{I\nu}(z_{2})=z_{12}^{-1}\dlt_{\mu\nu}+...,
\nmb
\al^{*}_{I\mu}(z_{1})\al_{I\nu}(z_{2})=z_{12}^{-1}\dlt_{\mu\nu}+...
\label{2.btgm1}
\enr

 An important property is the behavior of the $bc\bet\gm$ system under the local change of
coordinates on $A_{C_{I}}$ \cite{MSV}. For each new set of coordinates
\ber
b_{I\mu}=g_{\mu}(a_{I1},...,a_{Id}), \
a_{I\mu}=f_{\mu}(b_{I1},...,b_{Id})
\label{2.coordtr}
\enr
the isomorphic $bc\bet\gm$ system of fields is given by
\ber
b_{I\mu}(z)=g_{\mu}(a_{I1}(z),...,a_{Id}(z)),
\nmb
\bet_{I\mu}(z)={\d g_{\mu}\ov \d a_{I\nu}}(a_{I1}(z),...,a_{Id}(z))\al_{I\nu}(z),
\nmb
\bet^{*}_{I\mu}(z)={\d f_{\nu}\ov \d b_{I\mu}}(a_{I1}(z),...,a_{Id}(z))\al^{*}_{I\nu}(z),
\nmb
b^{*}_{I\mu}(z)={\d f_{\nu}\ov \d b_{I\mu}}(a_{I1}(z),...,a_{Id}(z))a^{*}_{I\nu}(z)+
\nmb
{\d^{2} f_{\lm}\ov\d b_{I\mu}\d b_{I\nu}}{\d g_{\nu}\ov\d a_{I\rho}}(a_{I1}(z),...,a_{Id}(z))\al^{*}_{I\lm}(z)\al_{I\rho}(z)
\label{2.coordtr1}
\enr 
Here the normal ordering of operators is implied. It is also understud, whenever necessary in what follows.

On the space $M_{C_{I}}$ the N=2 Virasoro superalgebra acts by the currents
\ber
G^{-}=\sum_{\mu}\al_{I\mu}a^{*}_{I\mu}, \
G^{+}=-a_{I0}\d\al^{*}_{I0}-\sum_{\mu\neq 0,I}\al^{*}_{I\mu}\d
a_{I\mu}, 
\
J=a_{I0}a^{*}_{I0}+\sum_{\mu\neq 0,I}\al^{*}_{I\mu}\al_{I\mu},
\nmb
T=\frac{1}{2}(a^{*}_{I0}\d a_{I0}-\d a^{*}_{I0}a_{I0})-\al_{I0}\d\al^{*}_{I0}+
\sum_{\mu\neq 0,I}(a^{*}_{I\mu}\d a_{I\mu}+\frac{1}{2}(\d\al^{*}_{I\mu}\al_{I\mu}-
\al^{*}_{I\mu}\d\al_{I\mu}))
\label{2.btgmvir}
\enr

 This algebra defines the mode expansion of the fields in Ramond sector
\ber
a_{I0}(z)=\sum_{n}a_{I0}[n]z^{-n-\frac{1}{2}}, \
a^{*}_{I0}(z)=\sum_{n}a^{*}_{I0}[n]z^{-n-\frac{1}{2}},
\nmb
\al_{I0}(z)=\sum_{n}\al_{I0}[n]z^{-n-1}, \
\al^{*}_{I0}(z)=\sum_{n}\al_{I0}[n]z^{-n},
\nmb
a_{I\mu}(z)=\sum_{n}a_{I\mu}[n]z^{-n}, \ 
a^{*}_{I\mu}(z)=\sum_{n}a^{*}_{I\mu}[n]z^{-n-1},
\nmb
\al_{I\mu}(z)=\sum_{n}\al_{I\mu}[n]z^{-n-\frac{1}{2}}, \
\al^{*}_{I\mu}(z)=\sum_{n}\al_{I\mu}[n]z^{-n-\frac{1}{2}},\ \mu\neq I
\label{2.btgmmod}
\enr
Then $M_{C_{I}}$ is generated by the creation operators acting on the Ramond vacuum state $|0>$
defined by the conditions 
\ber
a_{I\mu}[n]|0>=a^{*}_{I\mu}[n-1]|0>=\al_{I\mu}[n]|0>=\al^{*}_{I\mu}[n-1]|0>=0,
\ n>0.
\label{2.FockH}
\enr

 If the cone $C_{KJ}$ is a face of the cone $C_{K}$ and $C_{J}$ it is spanned by the
vectors $(e_{0},e_{1},...,\hat{e}_{K},...,\hat{e}_{J},...,e_{d})$.
We then consider the
BRST operator
\ber
D^{*}_{KJ}=Q^{*}_{0}+Q^{*}_{1}+...+\hat{Q^{*}}_{K}+...+\hat{Q^{*}}_{J}+...+Q^{*}_{d}
\label{2.DKJ}
\enr
acting on $\Phi_{C_{KJ}\oplus \Lm^{*}}$.
The space of sections $M_{C_{KJ}}$ of the chiral de Rham complex over the $A_{C_{KJ}}$ is given by the cohomology of $\Phi_{C_{KJ}\oplus \Lm^{*}}$ with respect to the operator $D^{*}_{KJ}$.
It is a localization of $M_{C_{K}}$ ($M_{C_{J}}$) with respect to the multiplicative system generated by $\prod_{\mu}(a_{K\mu}[0])^{m_{\mu}}$ ($\prod_{\mu}(a_{J\mu}[0])^{m_{\mu}}$), with $\sum_{\mu}m_{\mu}w^{*}_{K\mu}(C_{KJ})=0$ ($\sum_{\mu}m_{\mu}w^{*}_{J\mu}(C_{KJ})=0$).
Analogously, the localization maps can be defined for the cones which are the intersections of
an arbitrary number of maximal dimension cones 
~\cite{B}.

 The localization maps defined above allow to calculate the cohomology of the chiral de Rham
complex on $E$ as $\check{C}$ech cohomology of the covering by $A_{C_{I}}$, I=1,...,d ~\cite{B}:
\ber
0\rightarrow \oplus_{C_{I}}M_{C_{I}}\rightarrow \oplus_{C_{KJ}}M_{C_{KJ}}\rightarrow\cdots 
M_{C_{12...d}}\rightarrow 0
\label{2.Chech}
\enr
It finishes the calculation of $D_{\Delta}$-cohomology.

 The next step is to restrict the chiral de Rham complex on $E$ to the CY manifold 
$CY\subset  \mathbb{P}^{d-1}$. Let us define the function $W$ on $E$ which is linear on the fibers of $E$, so that in the coordinates $a_{I\mu}$ on $A_{C_{I}}$ this function is given by
\ber
W=a_{I0}(1+\sum_{\mu\neq 0,I}(a_{I\mu})^{d})
\label{2.W}
\enr
Then one has to introduce the corresponding screening currents and screening charges
\ber
S_{I0}(z)=\al_{I0}(z)(1+\sum_{\mu\neq 0,I}(a_{I\mu})^{d}(z)),
\nmb
S_{I\mu}(z)=\al_{I\mu}a_{I0}(z)(a_{I\mu})^{d-1}(z), \ \mu\neq 0,I,
\nmb
Q_{\mu}=\oint dz S_{I\mu}(z), \ \mu\neq I
\label{2.Wscren}
\enr
and the BRST operator 
\ber
D_{W}=\sum_{\mu\neq I}Q_{\mu}
\label{2.Wbrst}
\enr
(It is $D_{\Delta^{*}}$ differential in the notations of \cite{B}).
The cohomology of $M_{C_{I}}$ with respect to $D_{W}$ gives the space of sections $M_{C_{I}}|_{W}$ of chiral de Rham complex on $A_{C_{I}}\cap CY$ determined by the system of equations
\ber
a_{I0}=0,
\nmb
1+\sum_{\mu\neq 0,I}(a_{I\mu})^{d}=0.
\label{2.cyeq}
\enr

 The cohomology of the chiral de Rham complex on
$CY$ is calculated by $\check{C}$ech complex of the covering ~\cite{B}:
\ber
0\rightarrow \oplus_{C_{I}}M_{C_{I}}|_{W}\rightarrow \oplus_{C_{KJ}}M_{C_{KJ}}|_{W}\rightarrow \cdots
M_{C_{12...d}}|_{W}\rightarrow 0
\label{2.WChech}
\enr
Thus, we get $D_{\Delta}+D_{\Delta^{*}}$-cohomology.

 One can consider a more general function $W$ and  BRST operator (\ref{2.Wbrst}) adding
the monomial which corresponds to the internal point from the dual polytope $\Delta^{*}$, \cite{B}.

 It finishes the review of the chiral de Rham complex and its cohomology construction on CY hypersurface in $\mathbb{P}^{d-1}$.
 
\leftline {\bf 3. Line bundle twisted chiral de Rham complex.}

 In this section the generalization of Borisov's construction producing $O(N)$-twisted chiral de Rham complex on CY hypersurface is proposed.
 
 We twist the fermionic screening charges $Q^{*}_{i}$:
\ber
Q^{*}_{i}\rightarrow S^{*}_{i}[N_{i}]=\oint dz z^{N_{i}}S^{*}_{i}(z),\ N_{i}\in\mathbb{Z}
\label{3.twistchrg}
\enr
Then the old charges $Q^{*}_{i}$ can be considered zero modes of the screening
currents $S^{*}_{i}(z)=\sum_{n} S^{*}_{i}[n]z^{-n-1}$  and the BRST operator (\ref{2.dbrst})
is a particular case of more general one
\ber
D^{*}_{I}=S^{*}_{0}[N_{0}]+...+\hat{S^{*}}_{I}[N_{I}]+...+S^{*}_{d}[N_{d}]
\label{3.newdbrst}
\enr
Now the question is what are the cohomology of the space (\ref{2.iFock}) with respect to this new BRST operator?

 It follows by the direct calculation that the fields $a_{I\mu}(z)$, $\al_{I\mu}(z)$,
$\al^{*}_{I\mu}(z)$ from (\ref{2.btgm}) still commute with the new BRST operator (\ref{3.newdbrst}) but instead of $a^{*}_{I\mu}(z)$
we have to take
\ber
\nabla_{I\mu}(z)=a^{*}_{I\mu}(z)+N_{\mu}z^{-1}a^{-1}_{I\mu}(z)
\label{3.nabla}
\enr
The last term in this expression can be regarded as coming from $U(1)$ gauge potential on
$A_{C_{I}}$. We see in waht follows that this is indeed so and the modes of the fields $\nabla_{I\mu}(z)$ can be regarded as a string version of the covariant derivatives.

 In terms of this new $bc\bet\gm$ fields the $N=2$ Virasoro superalgebra currents are given by
\ber
G^{-}_{I}=\sum_{\mu}\al_{I\mu}\nabla _{I\mu}=\sum_{n}G^{-}_{I}[n]z^{-n-\frac{3}{2}}, 
\nmb
G^{+}_{I}=-a_{I0}\d\al^{*}_{I0}-\sum_{\mu\neq 0,I}\al^{*}_{I\mu}\d a_{I\mu}=
\sum _{n}G^{+}_{I}[n]z^{-n-\frac{3}{2}}, 
\nmb
J_{I}=a_{I0}\nabla _{I0}+\sum_{\mu\neq 0,I}\al^{*}_{I\mu}\al_{I\mu}=\sum_{n}J_{I}[n]z^{-n-1},
\nmb
T_{I}=\frac{1}{2}(\nabla _{I0}\d a_{I0}-\d \nabla _{I0}a_{I0})-\al_{I0}\d\al^{*}_{I0}+
\sum_{\mu\neq 0,I}(\nabla _{I\mu}\d a_{I\mu}+\frac{1}{2}(\d\al^{*}_{I\mu}\al_{I\mu}-
\al^{*}_{I\mu}\d\al_{I\mu}))=
\nmb
\sum_{n}L_{I}[n]z^{-n-2}
\label{3.nablavir}
\enr

To calculate the cohomology let us consider the vertex operator $V_{(0,p^{*})}(z)=
\exp{[p^{*}X]}(z)$, where $p^{*}\in \Lm^{*}$. We find
\ber
S^{*}_{\mu}[N_{\mu}](z_{1})V_{(0,p^{*})}(z_{2})=z_{12}^{N_{\mu}}S^{*}_{\mu}(z_{1})V_{(0,p^{*})}(z_{2})=
\nmb
z_{12}^{N_{\mu}+p^{*}(e_{\mu})}e_{\mu}\cdot\psi^{*}\exp [e_{\mu}\cdot X^{*}+p^{*}\cdot X](z_{2})+...,
\ \mu \neq I
\label{3.vacope}
\enr
Hence, the state $|(0,p^{*})>$ corresponding to the vertex $V_{(0,p^{*})}(0)$ is in $Ker (S^{*}_{\mu}[N_{\mu}])$ if $p^{*}(e_{\mu})\geq -N_{\mu}$. The (Ramond sector) state saturating the inequality is $|(0,-\sum_{\mu\neq I}N_{\mu}w^{*}_{I\mu})>$ and has the properties
\ber
\nabla _{I\mu}[k]|(0,-\sum_{\mu\neq I}N_{\mu}w^{*}_{I\mu})>=0, k\geq 0,
\nmb
a_{I\mu}[k]|(0,-\sum_{\mu\neq I}N_{\mu}w^{*}_{I\mu})>=
\nmb
\al_{I\mu}[k]|(0,-\sum_{\mu\neq I}N_{\mu}w^{*}_{I\mu})>=
\al^{*}_{I\mu}[k-1]|(0,-\sum_{\mu\neq I}N_{\mu}w^{*}_{I\mu})>=0,
\ k>0.
\label{3.truevac}
\enr
\leftline{\bf Proposition.}
The cohomology $\MM_{C_{I}}$ of $\Phi_{C_{I}\oplus \Lm^{*}}$ with respect to the differential
(\ref{3.newdbrst})
is generated from the vacuum state
\ber
|\Om_{I}>=|(0,-\sum_{\mu\neq I}N_{\mu}w^{*}_{I\mu})>
\label{2.Ivac}
\enr
by the creation operators of the fields (\ref{2.btgm}) and (\ref{3.nabla}).

 The proof is similar to the proof of Proposition 6.5. from \cite{B}.
 
 The vacuum $|\Om_{I}>$ defines the trivializing isomorphism 
of modules (over the chiral de Rham complex on $A_{C_{I}}$)
\ber
g_{I}: \MM_{C_{I}}\rightarrow M_{C_{I}}
\label{3.trivmap}
\enr
by the rule
\ber
g_{I}|\Om_{I}>=|0>,
\nmb
g_{I}(\nabla_{I\mu}[k])g_{I}^{-1}=a^{*}_{I\mu}[k], \
g_{I}(a_{I\mu}[k])g_{I}^{-1}=a_{I\mu}[k], 
\nmb
g_{I}(\al_{I\mu}[k])g_{I}^{-1}=\al_{I\mu}[k], \
g_{I}(\al^{*}_{I\mu}[k])g_{I}^{-1}=\al^{*}_{I\mu}[k]
\label{3.trivmap1}
\enr
We therefore call the vacuum $|\Om_{I}>$ the trivializing vacuum.

 Let us consider the subspace $\MM^{0}_{I}\subset \MM_{I}$ generated from $|\Om_{I}>$ by the operators $a_{I\mu}[0]$ and $\al_{I\mu}[0]$. The operator $G^{-}_{I}[0]$ acts on this subspace
by $\dlt_{I}=\sum_{\mu\neq I}\al_{I\mu}[0]\nabla_{I\mu}[0]$. It is natural to think that
$\MM^{0}_{I}$ is holomorphic de Rham complex over $A_{C_{I}}$ with coefficients in holomorphic line bundle. 

 On the intersections $A_{C_{I}}\cap A_{C_{J}}$ the relations
between the coordinates
\ber
a_{I0}=a_{J0}(a_{JI})^{d}, \ a_{I\mu}=a_{J\mu}a^{-1}_{JI}, \ \mu\neq I,J, \
a_{IJ}=a^{-1}_{JI}
\label{3.coordch}
\enr
can be used to find the relations between the trivializing vaccua
\ber
g_{I}|\Om_{I}>=\prod_{\mu\neq I}(a_{I\mu}[0])^{N_{\mu}}|\Om_{I}>=|0>,
\nmb
g^{-1}_{I}g_{J}|\Om_{J}>\equiv g_{IJ}|\Om_{J}>=|\Om_{I}>,
\nmb
g_{IJ}=(a_{(J)I}[0])^{N_{1}+N_{2}+...+N_{d}-dN_{0}}
\label{3.gijdef}
\enr
as well as between the sections
\ber
g_{IJ}:\MM^{0}_{J}\rightarrow \MM^{0}_{I}
\label{3.gij}
\enr
The functions $g_{IJ}$ from (\ref{3.gijdef})
are the transition functions
of the line bundle on $E$ which is induced from the $O(N)$-bundle on 
$\mathbb{P}^{d-1}$, where
\ber
N=N_{1}+...+N_{d}-dN_{0}.
\label{3.ON}
\enr
By this means, the set of modules $\MM^{0}_{C_{I}}$ with the differentials $\dlt_{I}$
and the transition functions (\ref{3.gijdef}) define the holomorphic de Rham complex
on $E$ with coefficients in the line bundle $\pi^{*}O(N)$. 

 One can extend this finite dimensional discussion to the infinite dimensional one.
We consider the relation between the currents $G^{-}_{I}(z)$ and  $G^{-}_{J}(z)$ on the
intersection $A_{C_{I}}\cap A_{C_{J}}$. Because of (\ref{3.coordch})
and (\ref{2.coordtr1}) we find
\ber
G^{-}_{I}(z)=G^{-}_{J}(z)+Nz^{-1}\al_{JI}(z)a^{-1}_{JI}(z)\Leftrightarrow
\nmb
G^{-}_{I}[k]=G^{-}_{J}[k]+N\sum_{m}\al_{JI}[m]a^{-1}_{JI}[k-m]
\label{3.coordchG}
\enr
In the finite-dimensional case the differentials $\dlt_{I}$ are consistent
on the intersections $A_{C_{I}}\cap A_{C_{J}}$: the difference $\dlt_{I}-\dlt_{J}$ coming
from the different trivializations is canceled by gauge transformation of the gauge potential: $A_{I\mu}=A_{J\mu}-g^{-1}_{IJ}\frac{\d g_{IJ}}{\d a_{J\mu}}$.  A similar event should occur
in the infinite-dimensional situation.   
Because the first Chern class on $E$ is zero, the second term in the expression (\ref{3.coordchG}) is due to different trivializations
defined on $A_{C_{I}}\cap A_{C_{J}}$ and has to be canceled by the gauge transformation
of the gauge potential:
\ber
G^{-}_{I}[k]=G^{-}_{J}[k]+N\sum_{m}\al_{JI}[m]a^{-1}_{JI}[k-m]-
\sum_{\nu \neq J}(g^{-1}_{IJ}\frac{\d g_{IJ}}{\d a_{J\nu}}\al_{J\nu})[k]=
G^{-}_{J}[k]
\label{3.gaugetr2}
\enr
Hence, the current $G^{-}\equiv G^{-}_{I}$ as well as the $N=2$ Virasoro superalgebra will
be globally defined if one takes into account the transformation of gauge potential and
extend the map (\ref{3.gij}) to the map
\ber
g_{IJ}(z)=(a_{(J)I}(z))^{N}: \MM_{C_{J}}\rightarrow \MM_{C_{I}}.
\label{3.gijext}
\enr
 
 If the cone $C_{IJ}$ is a face of the cone $C_{I}$ ($C_{J}$) and spanned by the
vectors $(e_{0},...,\hat{e}_{I},...,\hat{e}_{J},...e_{d})$, one can consider the
BRST operator 
\ber
D^{*}_{IJ}=S^{*}_{0}[N_{0}]+...+\hat{S^{*}}_{I}[N_{I}]+...+\hat{S^{*}}_{J}[N_{J}]+...
+S^{*}_{d}[N_{d}]
\label{3.brstIJ}
\enr
acting on $\Phi_{C_{IJ}\otimes \Lm^{*}}$.

%\leftline{\bf Proposition 2.} 
 The cohomology $\MM_{C_{IJ}}$ of $\Phi_{C_{IJ}\otimes \Lm^{*}}$ with respect to the differential (\ref{3.brstIJ}) is the localization of $\MM_{C_{I}}$ ($\MM_{C_{J}}$) with respect to the multiplicative system generated by $\prod_{\mu}(a_{I\mu}[0])^{m_{\mu}}$
($\prod_{\mu}(a_{J\mu}[0])^{m_{\mu}}$), with 
$\sum_{\mu}m_{\mu}w^{*}_{I\mu}(C_{IJ})=0$ ($\sum_{\mu}m_{\mu}w^{*}_{J\mu}(C_{IJ})=0$). 

 Thus, the module $\MM_{C_{IJ}}$ is generated from the vacuum vector
\ber
|\Om_{IJ}>=|(0,-\sum_{\mu\neq I,J}N_{\mu}w^{*}_{I\mu})>
\label{3.IJvac}
\enr
by the creation operators of the fields $a_{I\mu}(z)$, $\nabla_{I\mu}(z)$, $\mu\neq I,J$, $a_{IJ}(z)$, $a^{-1}_{IJ}(z)$, $a^{*}_{IJ}(z)$, $\al_{I\mu}(z)$, $\al^{*}_{I\mu}(z)$, $\mu\neq I$. $\MM_{C_{IJ}}$ can also be generated from the vacuum
\ber
|\tld{\Om}_{IJ}>=|(0,-\sum_{\mu\neq I,J}N_{\mu}w^{*}_{J\mu})>=
(a_{IJ}[0])^{N-N_{I}-N_{J}}|\Om_{IJ}>
\label{3.IJvac1}
\enr
by the creation operators of the fields $a_{J\mu}(z)$, $\nabla_{J\mu}(z)$, $\mu\neq I,J$, $a_{JI}(z)$, $a^{-1}_{JI}(z)$, $a^{*}_{JI}(z)$, $\al_{J\mu}(z)$, $\al^{*}_{J\mu}(z)$, $\mu\neq J$.

 Analogously, the modules $\MM_{C_{IJ...K}}$ and localization maps can be defined for the cones $C_{IJ...K}=C_{I}\cap C_{J}\cap...\cap C_{K}$. 
 
 Relation (\ref{3.IJvac1}) is a particular case of compatibility conditions localization maps to be satisfied for localization maps. They are as follows. For each maximal-dimension cone $C_{I}$ the trivializing vacuum $|\Om_{C_{I}}>$ defines a
linear function $\om^{*}_{I}\in\Lm^{*}$ on this cone:
\ber
\prod_{\mu\neq I}a^{-N_{\mu}}_{(I)\mu}(0)=\exp [-\om^{*}_{I}X](0), \
\om^{*}_{I}=dN_{0}w^{*}_{I0}+\sum_{\mu\neq 0,I}N_{\mu}w^{*}_{I\mu}.
\label{3.ordi}
\enr
It is easy to see that the collection of $\om^{*}_{I}$ satisfies the obvious compatibility condition. Namely, on the cone $C_{IJ}=C_{I}\cap C_{J}$ the functions $\om^{*}_{I}$
and $\om^{*}_{J}$ coincide and are given by the function $\om^{*}_{IJ}\in\Lm^{*}$ of the trivializing
vacuum $|\Om_{IJ}>$:
\ber
\prod_{\mu\neq I,J}a^{-N_{\mu}}_{(I)\mu}(0)=\exp [-\om^{*}_{IJ}X](0), \
\om^{*}_{IJ}=dN_{0}w^{*}_{I0}+\sum_{\mu\neq 0,I,J}N_{\mu}w^{*}_{I\mu}.
\label{3.ordij}
\enr
It can be verified that similar compatibility conditions are also satisfied for the functions $\om^{*}_{IJ...K}$ on the cones
$C_{IJ...K}=C_{I}\cap C_{J}\cap...\cap C_{K}$.  Then, the numbers $N_{0},...,N_{d}$
of screening currents modes define the support function $\om^{*}$ on $\Sgm$ \cite{Dan}, \cite{Ful}
of the toric divisor of the bundle $\pi^{*}O(N)$ on $E$. 

 Hence, similar to (\ref{2.Chech}) we have the $\check{C}$ech complex of the covering by $A_{C_{I}}$, $I=1,...,d$ 
\ber
0\rightarrow \oplus_{C_{I}}\MM_{C_{I}}\rightarrow \oplus_{C_{KJ}}\MM_{C_{KJ}}\rightarrow\cdots 
\MM_{C_{12...d}}\rightarrow 0
\label{3.Chech}
\enr
which gives the cohomology of chiral de Rham complex on $E$ twisted by 
$\pi^{*}O(N)$.

 The restriction of the twisted chiral de Rham complex on CY hypersurface is straightforward because BRST operator (\ref{2.Wbrst}) commutes with the operators (\ref{3.newdbrst}) and acts within each of the modules $\MM_{C_{IJ...K}}$. Therefore, the complex
\ber
0\rightarrow \oplus_{C_{I}}\MM_{C_{I}}|_{W}\rightarrow \oplus_{C_{KJ}}\MM_{C_{KJ}}|_{W}\rightarrow \cdots
\MM_{C_{12...d}}|_{W}\rightarrow 0
\label{3.WChech}
\enr 
gives the cohomology of $O(N)$-twisted chiral de Rham complex on CY hypersurface.
This completes the construction.

\leftline {\bf 4. The elliptic genus calculation.}

 In this section we calculate the elliptic genus of the twisted chiral de Rham complex closely following \cite{BL}.
The $q^{0}$ coefficient of the elliptic genus is related to the Hodge numbers of the sheaf
$O(N)$ on CY manifold. To justify the construction in Section 3, we calculate it for the case of torus $T^{2}\subset \mathbb{P}^{2}$ and $K3\subset \mathbb{P}^{3}$. 

 The discussion of Section 3 and the arguments of paper \cite{BL} allow extending the Definition
6.1. from \cite{BL} to the case under discussion: the elliptic genus is given by the supertrace
over the $\check{C}$ech cohomology space of the twisted chiral de Rham complex.

 The calculation is greatly simplified using the torus $(\mathbb{C}^{*})^{d}$ that acts on
$E$ \cite{BL}. We compute the function
\ber
\rho_{N}(CY,t_{1},...,t_{d},y,q)=
\sum_{k=1}^{d}(-1)^{k-1}\sum_{C_{I_{1}},...,C_{I_{k}}}superTr_{\MM_{C_{I_{1}...I_{k}}}}(\prod_{i=1}^{d}t^{K_{i}}_{i}y^{J[0]}q^{L[0]-\frac{c}{24}}),
\label{2.Elltor}
\enr
where $t_{i}$ are the formal variables grading the torus action with the help of generators
$K_{i}$ (whose explicit form is obvious) and then take the limit $t_{i}\rightarrow 1$, $i=1,...,d$ to get the elliptic genus $Ell_{N}(CY,y,q)$. 
It is quite helpful for subsequent computations 
to write the $N=2$ Virasoro superalgebra acting on $\MM_{C_{IJ...K}}$ in coordinates (\ref{2.ope})
\ber
G_{IJ...K}^{-}=-z^{-1}\om_{IJ...K}^{*}\cdot\psi-deg^{*}\cdot\d\psi+\psi\cdot\d X^{*},
\
G_{IJ...K}^{+}=-deg^{*}\cdot\d\psi^{*}+\psi^{*}\cdot\d X,
\nmb
J_{IJ...K}=-z^{-1}deg\cdot\om_{IJ...K}^{*}+deg\cdot\d X^{*}-deg^{*}\cdot\d X+\psi^{*}\cdot\psi,
\nmb
T_{IJ...K}=\frac{1}{2}(\d\psi^{*}\cdot\psi-\psi^{*}\cdot\d\psi)+
\d X\cdot(\d X^{*}-z^{-1}\om_{IJ...K}^{*})-\frac{deg}{2}\cdot\d (\d X^{*}-z^{-1}\om_{IJ...K}^{*})-
\nmb
\frac{deg^{*}}{2}\cdot\d^{2}X
\label{4.Virlog}
\enr
where
\ber
deg=e_{0}=\frac{1}{d}(e_{1}+...+e_{d}).
\label{4.degvect}
\enr
The supertraces over the modules $\MM_{C_{IJ...K}}$ can be calculated as the supertraces over the spaces $\Phi_{C_{IJ...K}\otimes \Lm^{*}}$
which are the complexes with respect to the differentials $D^{*}_{IJ...K}=S^{*}_{0}[N_{0}]+...+\hat{S^{*}}_{I}[N_{I}]+...+\hat{S^{*}}_{J}[N_{J}]+...+\hat{S^{*}}_{K}[N_{K}]+...+S^{*}_{d}[N_{d}]$. Because of (\ref{4.Virlog}) we obtain
\ber
\rho_{N}(CY,t_{1},...,t_{d},y,q)=
\nmb
y^{-\frac{d-2}{2}+d-1}\sum_{w^{*}\in \Lm^{*}}\prod_{i=1}^{d}t_{i}^{<w^{*},e_{i}>}
\sum_{C\subset\Sgm}(-1)^{codim C}\sum_{k\in C}
y^{-<deg^{*},k>+<w^{*}-\om^{*},deg>}q^{<w^{*}-\om^{*},k>}G(y^{-1},q)^{d}
\label{4.Elltor!}
\enr
where
\ber
G(y,q)=\prod_{k\geq 1}
\frac{(1-yq^{k-1})(1-y^{-1}q^{k})}
{(1-q^{k})^{2}}
\label{4.Gfunc}
\enr
and factor $y^{d-1}$ is caused by the Ramond vacuum.
This is a generalization of the elliptic genus expression obtained in \cite{BL}. It can be rewritten in terms of the theta functions. For that we need to use the trick
in \cite{BL} to get rid the positive-codimension cones contribution. Then we apply
"truly remarkable identity"
\ber
\prod_{k\geq 1}\frac{(1-tyq^{k-1})}{(1-tq^{k-1})}
\frac{(1-t^{-1}y^{-1}q^{k})}
{(1-t{-1}q^{k})}=
\sum_{n\in\mathbb{Z}}t^{n}(1-yq^{n})^{-1}G(y,q)
\label{A.Btremid}
\enr 
to write the maximal-dimension cones contribution as an infinite product. Thus,
we obtain
\ber
\rho_{N}(CY,t_{1},...,t_{d},y,q)=
y^{-\frac{d-2}{2}+d-1}\sum_{I=1}^{d}(\prod_{i=1}^{d}t_{i}^{-<\om^{*}_{I},e_{i}>})\frac{\Tt_{1,1}(t_{I}^{d},q)}{\Tt_{1,1}(t_{I}^{d}y,q)}
\prod_{J\neq I}\frac{\Tt_{1,1}(t_{I}^{-1}t_{J}y^{-1},q)}{\Tt_{1,1}(t_{I}^{-1}t_{J},q)},
\label{4.ellthet1}
\enr
where
\ber
\Tt_{1,1}(u,q)=q^{1/8}\prod_{n=0}(1-u^{-1}q^{n+1})(1-uq^{n})(1-q^{n+1})=
q^{1/8}\sum_{n\in Z}(-1)^{n}q^{(n^{2}-n)/2}u^{-n}
\nmb
\label{E.tt}
\enr

 In the limit $q\rightarrow 0$ $t_{i}\rightarrow 1$, the elliptic genus is related to the Hodge numbers
of the sheaf $O(N)$ on $CY$ hypersurface
\ber
Ell_{N}(CY,y)=y^{-\frac{d-2}{2}+d-1}\sum_{p,q}(-1)^{p+q}h^{p,q}(CY,O(N))y^{q}
\label{4.holEul}
\enr
We use this fact as a check of construction in two simplest examples, the torus $T^{2}$ in $\mathbb{P}^{2}$ and $K3$ in $\mathbb{P}^{3}$. 

 Taking the $t_{i}\rightarrow 1$ limit by l'Hopital's rule we find
\ber
Ell_{N}(T^{2},y)=3N(1-y)y^{-\frac{1}{2}},
\nmb
Ell_{N}(K3,y)=2((N^{2}+1)+(10-2N^{2})y+(N^{2}+1)y^{2})y^{-1}.
\label{4.T2K3}
\enr
We see that these expressions correctly reproduce the corresponding Hodge numbers.

\leftline {\bf 5. Concluding remarks.}

 In this note we presented generalization of Borisov's construction of chiral de Rham
complex on toric CY manifolds to include the CY hypersurfaces with
line bundles. It is shown that including non-zero modes of the screening currents
associated to the points of the polytope $\Delta$ of $\mathbb{P}^{d-1}$ into Borisov's differential  we
obtain $O(N)$-twisted chiral de Rham complex on the CY hypersurface. Moreover, we established the relation between the numbers of screening currents modes and the toric divisor support function for the line bundle $O(N)$ on $\mathbb{P}^{d-1}$. 

 We hope that the construction discussed above can be applied for the quantization of monad bundles in the heterotic string models. Another possible application appears if we consider
the twisting line bundle as a Chan-Paton bundle of a bound state of $(2d-4,2d-6)$ $D$-branes on the CY. In this context it would be interesting to generalize the construction to include also Chan-Paton sheafs describing more general bound states of the $D$-branes on CY \cite{HM}. 
 
 There are two more questions to be mentioned. 
The first question which is obvious is to extend the discussion
for CY hypersurfaces in general toric manifolds. The second one is a possible mirror symmetry generalization. In the construction of Borisov
the differentials associated to the pair of reflexive polytopes $\Delta$ and $\Delta^{*}$
come into play on the equal footing which makes the mirror symmetry explicit \cite{B},
\cite{BL}. For the generalization considered in this paper these democracy seems to be
broken. Indeed, if one first takes the cohomology with respect to the differential $D_{\Delta^{*}}$ which is unchanged, we obtain the usual (untwisted) chiral de
Rham complex on toric manifold $\mathbb{P}_{\Delta^{*}}$. It is difficult to believe
that taking then the cohomology
with respect to the generalized differential $D_{\Delta}$ as a second step we restrict the chiral de Rham complex to a mirror CY hypersurface in $\mathbb{P}_{\Delta^{*}}$. Therefore, the question is how to extend the mirror symmetry to this case. The more general setup is simultaneous generalization of differentials $D_{\Delta}$ and $D_{\Delta^{*}}$ by the nonzero screening currents modes.

  Recently, paper \cite{BK} by L.Borisov and R.Kaufmann has appeared where a different construction of the chiral de Rham complex twisted by a vector bundle has been presented. It would be interesting to understand the relation with our approach.

\leftline{\bf Acknowledgments.}

I thank to B.L.Feigin for the helpful discussions on the cohomology of nonzero mode of fermionic screening current. I am especially gratefull to
L.Borisov for explanations on papers \cite{B}, \cite{BL}. This work is supported 
within the framework of the federal 
program "Scientific and Scientific-Pedagogical Personnel of Innovational Russia" (2009-2013), by the RFBR initiative interdisciplinary project (Grant No. 09-02-
12446-ofi-m) as well as by grants RFBR-07-02-00799-a, SS3472.2008.2,
RFBR-CNRS 09-02-93106.

\end{document}